\documentclass{jpsj2}

\title{
Mechanism of Ambipolar Field-Effect Carrier Injections in One-Dimensional Mott Insulators
}

\author{
Kenji \textsc{Yonemitsu}$^{1,2}$
\thanks{E-mail address: kxy@ims.ac.jp}
}

\inst{
$^1$Institute for Molecular Science, Okazaki 444-8585 \\
$^2$Graduate University for Advanced Studies, Okazaki 444-8585
}

\abst{
To clarify the mechanism of recently reported, ambipolar carrier injections into quasi-one-dimensional Mott insulators on which field-effect transistors are fabricated, we employ the one-dimensional Hubbard model attached to a tight-binding model for source and drain electrodes. To take account of the formation of Schottky barriers, we add scalar and vector potentials, which satisfy the Poisson equation with boundary values depending on the drain voltage, the gate bias, and the work-function difference. The current-voltage characteristics are obtained by solving the time-dependent Schr\"odinger equation in the unrestricted Hartree-Fock approximation. Its validity is discussed with the help of the Lanczos method applied to small systems. We find generally ambipolar carrier injections in Mott insulators even if the work function of the crystal is quite different from that of the electrodes. They result from balancing the correlation effect with the barrier effect. For the gate-bias polarity with higher Schottky barriers, the correlation effect is weakened accordingly, owing to collective transport in the one-dimensional correlated electron systems.
}

\kword{
ambipolar field-effect transistor, organic Mott insulator, Schottky barrier, one-dimensional Hubbard model, time-dependent Schr\"odinger equation, Poisson equation
}

\begin{document}
\maketitle

\section{Introduction} 

Molecular materials have been extensively investigated so far. Intriguing properties have been observed recently in non-equilibrium conditions. To seek new functions of molecular materials that reflect collective characters in non-equilibrium conditions, we here study the field-effect properties of one-dimensional Mott insulators. 

Among the electronic devices, the metal-oxide-semiconductor field-effect transistor (MOSFET) is the most important for very-large-scale integrated circuits. Its current-voltage characteristics are often explained rather simply. One of the conditions that allow the simple explanation is the so-called gradual channel approximation, which requires the fact that the transverse field (perpendicular to the insulating layer) in the channel is much larger than the longitudinal field (parallel to the channel). \cite{Sze_book} The total charge per unit area induced in the semiconductor is given by the potential difference between the gate electrode and the respective position along the channel, which is multiplied by the gate insulator capacitance per unit area. The conductivity is then given by the charge density multiplied by the mobility, which is usually regarded as a constant. 

Some field-effect transistors fabricated on organic semiconductors/insulators have very different situations from the above. For example, carbon nanotube field-effect transistors operate as unconventional Schottky barrier transistors, in which transistor action occurs primarily by varying the contact resistance rather than the channel conductance. \cite{Heinze_PRL,Appen_PRL02,Rado_APL03,Javey_Nature} Although the gate voltage was initially assumed to modify the nanotube conductance in analogy with the channel of an ordinary field-effect transistor (operation involves gate modulation of carrier density in the channel), Schottky barriers at the contacts are now known to play a central role (operation involves gate modulation of the Schottky barrier height/width).  The Schottky barriers are sensitive to the work-function difference between the channel and the source/drain electrodes. By matching their work functions (reducing the work function of the electrode by exposing it to hydrogen \cite{Javey_Nature} or increasing it by oxygen exposure \cite{Heinze_PRL}), the ambipolar field-effect characteristics are achieved. The one-dimensional structure and the coherent transport in carbon nanotubes make the contact resistance relatively important, so that the conductivity is no more given by the charge density multiplied by the mobility. 

Many molecular materials have strong anisotropy in electronic conduction owing to their molecular arrangements in the crystals. In addition, electron correlations also contribute to the confinement of the motion of carriers. \cite{Kishine_IJMPB} Recently, Hasegawa {\it et al.} reported ambipolar field-effect characteristics in metal-insulator-semiconductor field-effect transistor (MISFET) device structures based on organic single crystals of the quasi-one-dimensional Mott insulator (BEDT-TTF)(F$_2$TCNQ) [BEDT-TTF=bis(ethylenedithio)tetrathiafulvalene, F$_2$TCNQ=2,5-difluorotetracyanoquinodimethane]. \cite{Hasegawa_PRB04} Here again, the Schottky barriers play an important role owing to the molecular arrangement. In the crystals, the molecular long axes are perpendicular to the gate insulator. The direction of the largest intermolecular overlap, which corresponds to a transfer integral of about 0.2eV between the neighboring BEDT-TTF molecules, is parallel to that of the drain current. \cite{Hasegawa_PRB00} The transfer integral between the neighboring BEDT-TTF and F$_2$TCNQ molecules parallel to the gate insulator is about 0.01eV, \cite{Hasegawa_PRB00} while the transfer integrals along the molecular long axes are much smaller. Thus the motion of carriers is quasi-one-dimensional, and especially the motion perpendicular to the gate insulator rarely takes place. 

In contrast to the situation involving band insulators, the effect of the interface barrier potentials between the Mott insulator and the source/drain electrodes on electron injections is found very similar to that on hole injections at the source and drain contacts. \cite{Hasegawa_PRB04} Namely, the injections of electrons and holes are almost equally possible down to low temperatures. The work function of the source/drain electrodes is not needed to match with that of the crystal if it is generally realized. In this paper, we theoretically show that the ambipolar field-effect characteristics generally result from combined effects of the electron correlation and the Schottky barriers. We calculate the current-voltage characteristics using the one-dimensional Hubbard model for Mott insulators (and a tight-binding model with alternating transfer integrals, alternating site energies, or staggered magnetic fields for band insulators) and adding to it scalar and vector potentials that originate from the long-range Coulomb interaction and are modified by the work-function difference, the gate bias, and the drain voltage. The time-dependent Schr\"odinger equation is combined with the Poisson equation and numerically solved self-consistently at each site and time within the unrestricted Hartree-Fock approximation.

The ambipolar field-effect characteristics are realized in a wide parameter space spanned by the work-function and bandwidth differences between the crystal and the electrodes as well as the strength of the long-range Coulomb interaction governing the potential distribution. This is shown to be achieved by balancing the correlation effect with the barrier effect. For the gate-bias polarity with higher Schottky barriers, the correlation effect is weakened accordingly. The charge is transported in a very collective manner in one-dimensional Mott insulators in strong-enough electric fields. Such collective properties are reminiscent of the knock-on mechanism in ion channels. \cite{Roux_Nature,Roux_PNAS}

\section{One-Dimensional Model for a Crystal Attached to Electrodes}

Because the carrier motion perpendicular to the gate insulator rarely takes place owing to the negligible intermolecular overlaps along the molecular long axes, we assume that the effect of the gate electrode appears only in a boundary value of the Poisson equation for the source-drain potential. This ``boundary'' is actually set at the middle of the crystal below. Electrons or holes are injected at the source and drain electrodes. Because they move along the crystal and these electrodes, we need to include them explicitly in the model. 

We use the one-dimensional Hubbard model for Mott insulators attached to the source and drain electrodes represented by a tight-binding model. The total number of electrons is the same as the number of sites. In general, the work function of the crystal is different from that of the source and drain electrodes. The work-function difference $ \phi $ is so defined that $ \phi > 0 $ if the crystal has a higher work function than the electrodes. The model is written as
\begin{eqnarray}
H & = & \sum_i \left( \epsilon_i + v_i \right) n_i    \nonumber \\
& & - \sum_{i,\sigma} \left[ t_{i,i+1}(t) c_{i,\sigma}^\dagger c_{i+1,\sigma} 
+ t_{i+1.i}(t) c_{i+1,\sigma}^\dagger c_{i,\sigma} \right]    \nonumber \\
& & + \sum_i U_i \left( n_{i\uparrow} - 1/2 \right) 
\left( n_{i\downarrow} - 1/2 \right)
\;, \label{eq:Hamiltonian}
\end{eqnarray}
where $ c_{i,\sigma}^\dagger $ ($ c_{i,\sigma} $) creates (annihilates) an electron with spin $ \sigma $ at site $ i $, $ n_{i\sigma} = c_{i,\sigma}^\dagger c_{i,\sigma} $, and $ n_i = \sum_\sigma n_{i\sigma} $. The site energy $ \epsilon_i $ is set at $ -\phi $ in the crystal and at $ 0 $ in the electrodes. The absolute value of the transfer integral $ \mid t_{i,i+1}(t) \mid $ is set at $ t_{\rm c} $ if either $ i $ or $ i + 1 $ is in the crystal and at $ t_{\rm e} $ otherwise. The on-site repulsion $ U_i $ is set at $ U $ in the crystal and at $ 0 $ in the electrodes. 

Later, we consider band insulators, for comparison, by using a tight-binding model ($ U $=0) with alternating transfer integrals [$ t_{\rm c} $ above is replaced by $ t_{\rm c} - (-1)^{i} \delta t $], alternating site energies [$ - \phi $ above is replaced by $ - \phi + (-1)^{i} \delta \epsilon $], or staggered magnetic fields [$ - \phi $ above is replaced by $ - \phi + (-1)^{i} h $ for up-spin and $ - \phi - (-1)^{i} h $ for down-spin electrons].

Hereafter, the drain voltage (relative to the source) is denoted by $ V_{\rm D} $, and the gate voltage (relative to the source) by $ V_{\rm G} $. Following ref.~\citen{Hasegawa_PRB04}, we adopt the symmetric-gate operation, in which the electric reference point of the new gate voltage $ U_{\rm G} $ is set at the middle of the source-drain potential, $ U_{\rm G} = V_{\rm G} - V_{\rm D}/2 $. Even if the external field is not applied, $ V_{\rm D} = U_{\rm G} = 0 $, electrons are so transferred from the low work-function side to the high work-function side that the chemical potentials coincide to reach equilibrium. There appears a scalar potential $ v_i $ from the redistributed charge density. Note that the scalar potential in electromagnetism is given by dividing it by $ -e $. In order to use the periodic boundary condition for finite $ V_{\rm D} $, we introduce a time-dependent vector potential $ A(t) $ along the channel. Then the transfer integral acquires the Peierls phase, 
\begin{equation}
t_{i,i+1}(t) = t_{i+1,i}^\ast(t) = \mid t_{i,i+1}(t) \mid 
\exp \left[ {\rm i} \frac{ea}{\hbar c} A(t) \right]
\;,
\end{equation}
where $ t $ denotes time, $ e $ the absolute value of the electronic charge, $ a $ the lattice constant, and $ c $ the light velocity. In order for the Poisson equation to contain only the scalar potential, $ v_i $, we use the space-independent vector potential,
\begin{equation}
A(t) = -c \int^t {\rm d}t' E_D
\;,
\end{equation}
for the space-independent component of the electric field, 
\begin{equation}
E_D = - V_{\rm D} / L 
\;,
\end{equation}
with $ L $ being the number of sites. The gauge invariance of the motion of electrons is proved in Appendix.

The scalar potential $ v_i $ obeys the Poisson equation on the discrete lattice,
\begin{equation}
v_{i+1} - 2 v_i + v_{i-1} = - V_{{\rm P}i} \left( \langle n_i \rangle - n_{{\rm B}i} \right)
\;,
\end{equation}
where the parameter $ V_{{\rm P}i} $ comes from the long-range Coulomb interaction and $ n_{{\rm B}i} $ from the background charge. Note that the Poisson equation itself is for three dimensions, but the scalar potential is assumed to depend on the position along the channel. If we take an orthorhombic system with lattice constants $ a $, $ b $, and $ c $ (not the light velocity here) and dielectric permittivity $ \epsilon $ (not the site energy here), the Gauss law gives $ V_{{\rm P}i} = 4\pi a^2/(bc) \times e^2/(4\pi \epsilon a) $, which is proportional to the so-called nearest-neighbor repulsion strength and inversely proportional to the dielectric permittivity $ \epsilon $. The parameter $ V_{{\rm P}i} $ is set at $ V_{{\rm Pc}} $ in the crystal and at $ V_{{\rm Pe}} $ in the electrodes, with $ V_{{\rm Pe}} \ll V_{{\rm Pc}} $. In the ideally metallic electrodes, $ V_{{\rm Pe}} $ would be zero. We use a small but finite value for it to stabilize numerical calculations. For the background charge, the neutrality condition, 
\begin{equation} 
\sum_i \langle n_i \rangle = \sum_i n_{{\rm B}i} 
\;,
\end{equation} 
is crucial for the boundary conditions at the crystal-electrode interfaces to be compatible with those far from the interfaces. For $ V_{\rm D} $=0, the parameter $ n_{{\rm B}i} $ is set at $ n_{{\rm Bc}} $ uniformly inside the crystal and at $ n_{{\rm Be}} $ uniformly in the electrodes. They depend on the gate bias $ U_{\rm G} $ and the drain voltage $ V_{\rm D} $. Concerning the site dependence of $ n_{{\rm B}i} $ for $ V_{\rm D} \neq 0 $, we have compared the results with different assumptions: i) $ n_{{\rm B}i} $ in the crystal is a constant in the left half and another constant in the right half; and ii) $ n_{{\rm B}i} $ in the crystal is a linear function of $ i $. In both cases, the site dependence of $ n_{{\rm B}i} $ in the electrodes is also set like that in the crystal. With either of these assumptions, $ n_{{\rm B}i} $ is numerically obtained with a solution $ v_i $ to the Poisson equation by imposing the self-consistency. These assumptions are found to produce quantitatively similar results. To show the data below, we use the assumption i).

\begin{figure}
\includegraphics[height=6cm]{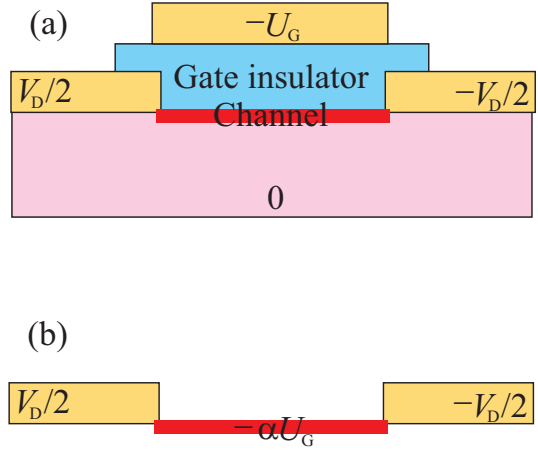}
\caption{(a) Device geometry, with source and drain electrodes on the left and right, respectively, and a gate electrode on top. (b) One-dimensional model, where the effect of the gate bias is included in a boundary value of the Poisson equation.}
\label{fig:fet}
\end{figure}
In principle, the gate bias $ U_{\rm G} $ directly determines the potential of the gate electrode, not of the channel. The potential distribution in the gate insulator is determined by the Laplace equation, with boundary conditions set by the electrodes, which are coupled with the Poisson equation for the channel [Fig.~\ref{fig:fet}(a)]. For a solution to the Laplace equation for the gate insulator, see for example ref.~\citen{Heinze_PRL}. The gate bias $ U_{\rm G} $ consequently modulates the potential along the channel. For small $ \mid U_{\rm G} \mid $, the modulation of the potential at the midpoint would be linear with respect to $ U_{\rm G} $. Even if the modulation contains higher-order terms, they would not alter the conclusion below on whether the field-effect characteristics are ambipolar or unipolar. Then we phenomenologically introduce the gate efficiency factor $ \alpha $ \cite{Javey_APL02} [Fig.~\ref{fig:fet}(b)] instead of considering the transverse field. The details of the gate insulator including its shape and size are thus ignored. When the voltage at the gate electrode is varied in some range, the amount by which the potential changes at the midpoint of the channel is smaller than this range. Thus $ \alpha $ is generally smaller than unity. For the thinner dielectric film (i.e., for the larger gate capacitance), the smaller gate-bias modulation causes a given amount of the drain-current modulation. \cite{Appen_PRL02,Rado_APL03} This can be interpreted as due to the larger $ \alpha $.

In such circumstances as described above, the gate bias $ U_{\rm G} $ appears with the gate efficiency factor $ \alpha $ in a boundary value of the Poisson equation, 
\begin{equation}
v_{L/2} - v_0 = - \alpha U_{\rm G} + \phi
\;,
\end{equation}
where $ i = L/2 $ is set at the middle of the crystal and $ i = 0 $ in the electrode is the furthest point from the crystal-electrode interfaces. Because $ U_{\rm G} $ appears only here as a boundary value, we rewrite $ \alpha U_{\rm G} $ as $ U_{\rm G} $ for simplicity. Then the boundary condition is given by 
\begin{equation}
v_{L/2} - v_0 = - U_{\rm G} + \phi
\;.
\end{equation}
When the vector potential is added to the scalar potential, the total potential is given by $ v_i - V_{\rm D} (i/L-1/2) $, which becomes $ + V_{\rm D}/2 $ deep inside the source electrode ($ i = 0 $), $ - U_{\rm G} + \phi $ at the middle of the crystal ($ i = L/2 $), and $ - V_{\rm D}/2 $ deep inside the drain electrode ($ i = L $) if $ v_0 = v_L $ is arbitrarily set at zero. 

For each value of $ U_{\rm G} $, we first set $ V_{\rm D} $ at zero and iteratively solve the eigenvalue equation for the ground state in the unrestricted Hartree-Fock approximation simultaneously with the Poisson equation for the potential distribution. After setting $ V_{\rm D} $ at a finite value, we solve the time-dependent Schr\"odinger equation in the same approximation by decomposing the exponential operator, as performed in ref.~\citen{Miyashita_JPSJ03}. The self-consistency is imposed at each site and time both in the time-dependent Schr\"odinger equation and in the Poisson equation. Thus the potential distribution is obtained simultaneously with the current density in the non-equilibrium condition. 

The drain current $ I_{\rm D} $ is defined in the same manner as in ref.~\citen{Oka_PRL03},
\begin{equation}
I_{\rm D} = \frac{1}{\Delta t} \int_{0}^{\Delta t} {\rm d}t \langle J(t) \rangle
\;,
\end{equation}
where the current density at time $ t $, given by
\begin{equation}
J(t) = - \frac{1}{L} \sum_{i,\sigma} \left[ 
{\rm i} t_{i+1.i}(t) c_{i+1,\sigma}^\dagger c_{i,\sigma}
- {\rm i} t_{i,i+1}(t) c_{i,\sigma}^\dagger c_{i+1,\sigma} 
\right]
\;,
\end{equation}
is averaged over the period, $ 0 < t < \Delta t $ with $ \Delta t = 2\pi L / ( 4 V_{\rm D}) $. Its sign is set so that generally $ I_{\rm D} >(<) 0 $ for $ V_{\rm D} >(<) 0 $. The time-evolution of the current density was compared with the exact one \cite{Oka_PRL03} for the pure Hubbard model (without electrodes) at and near half filling in a wide range of $ V_{\rm D} $ with the periodic boundary condition. The results are quite similar, sharing the qualitative characters. It is like teeth of a saw as a function of time for small $ V_{\rm D} $, while it is a sinusoidal function with a large amplitude for large $ V_{\rm D} $. Later we compare the drain current obtained by the Lanczos method and that obtained in the unrestricted Hartree-Fock approximation for small systems attached to electrodes. Both results show ambipolar field-effect characteristics in Mott insulators, but unipolar ones in band insulators.

\section{Results}

In the quasi-one-dimensional Mott insulator (BEDT-TTF)(F$_2$TCNQ) used for the field-effect transistors, the intermolecular overlap along the channel is estimated at 0.019, \cite{Hasegawa_PRB00} which corresponds to a transfer integral of about 0.2eV, while the polarized reflectivity spectra show a charge gap of about 0.7eV. \cite{Hasegawa_SSC97} The work-function difference is estimated at $ \phi \sim -1$eV with gold electrodes. \cite{Hasegawa_PRB04} For numerical calculations, we use $ t_{\rm c} $=1, as a unit, and $ U $=2 for Mott insulators, which produces a charge gap of about $ \Delta_{\rm CG} $=0.68 at half filling in the present approximation. In most of the calculations below, we use $ t_{\rm e} $=1. Thus, compared with the experimental values, the transfer integral in the crystal is larger, while that in the electrodes is smaller. We use them simply because the numerical convergence is generally stable and rapid for comparable transfer integrals. As shown later, the main conclusion is not altered by different transfer integrals. Below we frequently use $ V_{{\rm Pc}} $=0.05 and $ V_{{\rm Pe}} $=10$^{-3}$, so that $ V_{{\rm Pc}} / V_{{\rm Pe}} $=50. The numerical results are insensitive to $ V_{{\rm Pc}} $ or $ V_{{\rm Pe}} $ if the ratio is unchanged, so that we mention the ratio only. It is shown later that different ratios do not alter the main conclusion. Below we mostly use the 100-site periodic system with the central 51 sites belonging to the crystal and the peripheral 49 sites to the electrodes.

\begin{figure}
\includegraphics[height=12cm]{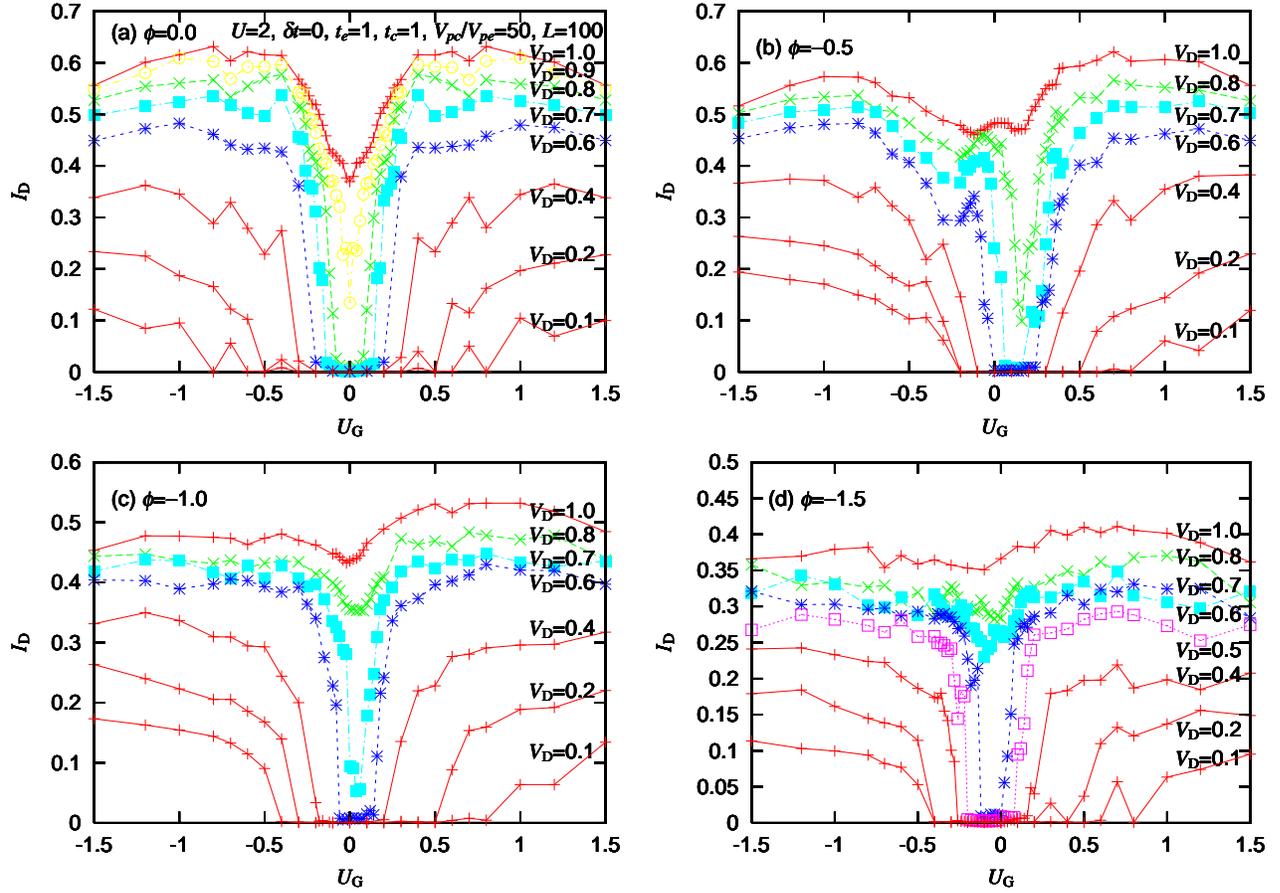}
\caption{$ I_{\rm D} $-$ U_{\rm G} $ characteristics at various drain voltage $ V_{\rm D} $ for Mott insulators with $ U $=2 and $ \delta t $=0. The work-function differences are, (a) $ \phi $=0, (b) $ \phi $=$-$0.5, (c) $ \phi $=$-$1.0, and (d) $ \phi $=$-$1.5. The other parameters are $ t_{\rm e} $=$ t_{\rm c} $=1, $ V_{{\rm Pc}} / V_{{\rm Pe}} $=50, and $ L $=100.}
\label{fig:Mott_IV}
\end{figure}

\begin{figure}
\includegraphics[height=12cm]{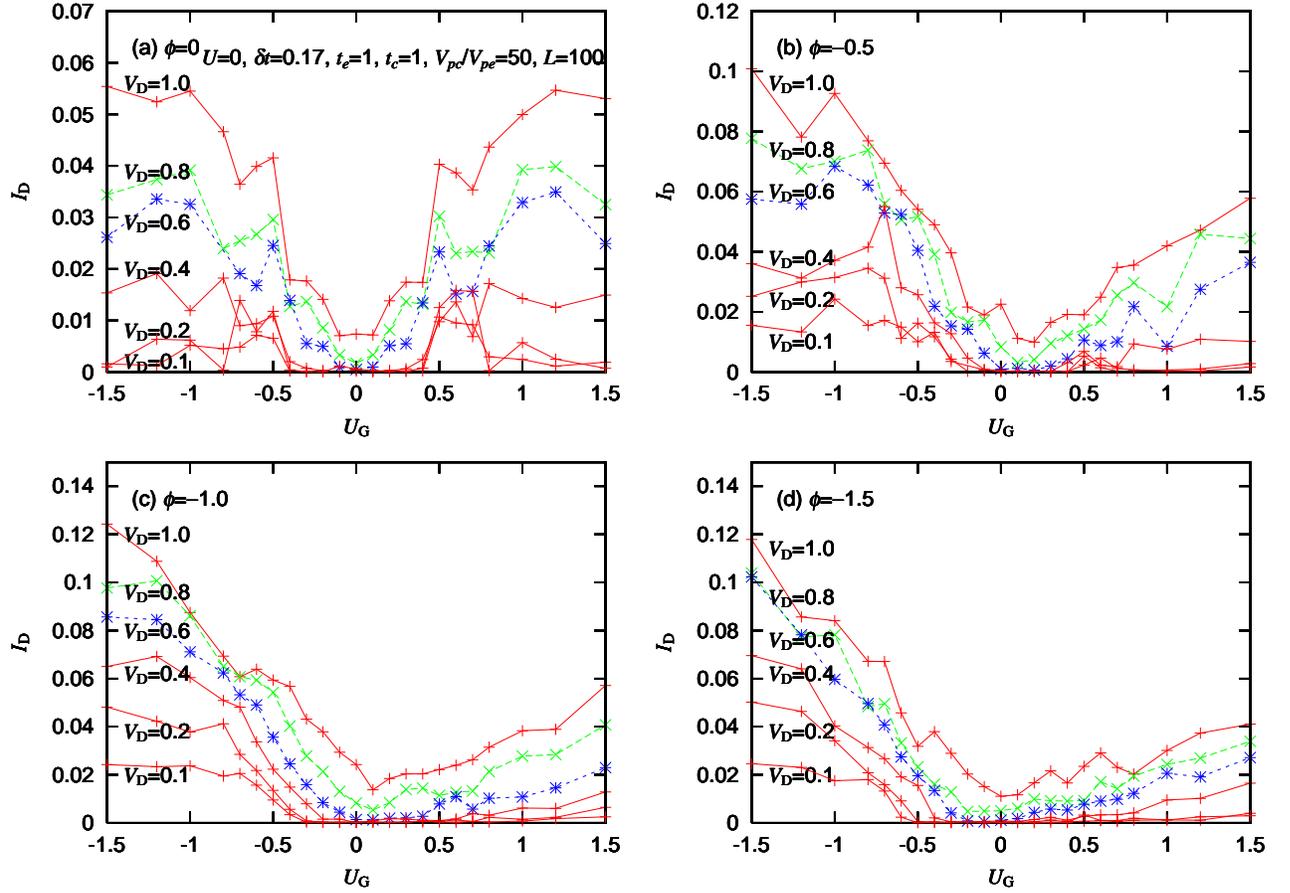}
\caption{$ I_{\rm D} $-$ U_{\rm G} $ characteristics at various drain voltage $ V_{\rm D} $ for band insulators with $ U $=0 and $ \delta t $=0.17. The work-function differences are, (a) $ \phi $=0, (b) $ \phi $=$-$0.5, (c) $ \phi $=$-$1.0, and (d) $ \phi $=$-$1.5. The other parameters are the same as in Fig.~\ref{fig:Mott_IV}.}
\label{fig:band_IV}
\end{figure}

When the work function of the crystal and that of the electrodes coincide ($ \phi $=0), the drain-current ($ I_{\rm D} $)-gate-voltage ($ U_{\rm G} $) characteristics are symmetric with respect to the exchange of $ U_{\rm G} $ and $ - U_{\rm G} $ owing to the electron-hole symmetry. This property is irrespective of whether the electronic state is a Mott insulator [Fig.~\ref{fig:Mott_IV}(a)] or a band insulator [Fig.~\ref{fig:band_IV}(a)].

Mott insulators show nearly symmetric $ I_{\rm D} $-$ U_{\rm G} $ characteristics for finite work-function differences (Fig.~\ref{fig:Mott_IV}). For small $ V_{\rm D} $ and $ \mid U_{\rm G} \mid < \Delta_{\rm CG}/2 $, $ I_{\rm D} $ is suppressed by the charge gap. The drain current $ I_{\rm D} $ increases with electron injections ($ U_{\rm G} > 0 $) and with hole injections ($ U_{\rm G} < 0 $) in a similar manner for any $ V_{\rm D} $. For large $ V_{\rm D} $, further carrier injections (with increasing $ \mid U_{\rm G} \mid $) lead $ I_{\rm D} $ to a maximum and slight reduction after reaching the maximum. This is due to the small bandwidth of the electrodes used here. In fact, this behavior disappears for electrodes with a wide bandwidth, as shown later. It is not reported in the experiment with gold electrodes as well. \cite{Hasegawa_PRB04} 

Around $ U_{\rm G} $=0, a threshold exists in $ V_{\rm D} $ above which $ I_{\rm D} $ becomes finite and rapidly increases with $ V_{\rm D} $ for Mott insulators. The threshold $ V_{\rm Dth} $ is in the range 
$ 0.8 < V_{\rm Dth} < 0.9 $ for $ \phi = 0 $ [Fig.~\ref{fig:Mott_IV}(a)], 
$ 0.7 < V_{\rm Dth} < 0.8 $ for $ \phi = -0.5 $ [Fig.~\ref{fig:Mott_IV}(b)], 
$ 0.6 < V_{\rm Dth} < 0.7 $ for $ \phi = -1.0 $ and $ -1.5 $ [Figs.~\ref{fig:Mott_IV}(c) and \ref{fig:Mott_IV}(d)], and 
$ 0.5 < V_{\rm Dth} < 0.6 $ for $ \phi = -2.0 $ (not shown).
The 120-site system with 63 crystal sites and 57 electrode sites has slightly different thresholds, e.g., $ 0.7 < V_{\rm Dth} < 0.8 $ for $ \phi = -1.0 $ (not shown), but the tendency is unchanged.
The threshold $ V_{\rm Dth} $ moderately decreases with increasing work-function difference. Above the threshold $ V_{\rm Dth} $, the $ I_{\rm D} $-$ U_{\rm G} $ characteristics generally deviate from the symmetric ones largely compared with those below $ V_{\rm Dth} $. There appears no general rule concerning the asymmetry (e.g., which of the electron injections and the hole injections are more effective in increasing $ I_{\rm D} $) above $ V_{\rm Dth} $. The asymmetry depends on the bandwidth of the electrodes, as shown later. The largest $ I_{\rm D} $ value at the optimum $ U_{\rm G} $ at a given $ V_{\rm D} > V_{\rm Dth} $ moderately decreases with increasing work-function difference (see the vertical scales). In general, the larger work-function differences bring about the energetically higher and spatially wider Schottky barriers, so that the charge gap is more easily collapsed by the drain voltage (thus $ V_{\rm Dth} $ is reduced) and the drain current is generally reduced.

Below $ V_{\rm Dth} $, however, we find a general tendency (as shown here and below with different parameters) concerning the deviations from the symmetric $ I_{\rm D} $-$ U_{\rm G} $ characteristics. When the work function of the crystal is lower than that of the electrodes ($ \phi < 0 $), $ I_{\rm D} $ is slightly larger for $ U_{\rm G}  < 0 $ than for $ U_{\rm G} > 0 $ for a given $ \mid U_{\rm G} \mid $ at $ V_{\rm D} < V_{\rm Dth} $, as if the hole mobility is higher than the electron mobility. This is consistent with the experimental finding. \cite{Hasegawa_PRB04} It can be concluded from this result with the electron-hole exchange operation that, for $ \phi > 0 $, the $ I_{\rm D} $-$ U_{\rm G} $ characteristics behave as if the electron mobility is higher. The drain current $ I_{\rm D} $ sometimes shows an oscillating dependence on $ U_{\rm G} $ especially for small $ V_{\rm D} $. It is caused partly by averaging the current density, which generally oscillates with a $ U_{\rm G} $-dependent frequency, \cite{Oka_PRL03} over the fixed time period, and partly by iteratively obtaining the potential distribution during the application of the drain voltage. It does not have any physical meaning.

The drain current $ I_{\rm D} $ has a minimum at $ U_{\rm G} = U_{\rm G}^{\rm min} $. The minimum point $ U_{\rm G}^{\rm min} $ is exactly zero for $ \phi $=0, but it generally deviates from zero. The dependence of $ U_{\rm G}^{\rm min} $ on $ \phi $ is not monotonic. It is shifted to a positive value for small $ \mid \phi \mid $ and then to a negative value for large $ \mid \phi \mid $ in Fig.~\ref{fig:Mott_IV} for $ \phi < 0 $. The sign of $ U_{\rm G}^{\rm min} $ can be altered by using a different bandwidth of the electrodes, as shown later. Thus the behavior of $ U_{\rm G}^{\rm min} $ is not simple. In the experiment, $ U_{\rm G}^{\rm min} $ is reported to be 2-3V. \cite{Hasegawa_PRB04} If the gate efficiency factor were unity, it would be much smaller.

We have a few comments on the experimental observation. The nonlinearity in $ I_{\rm D} (U_{\rm G} \neq 0) - I_{\rm D}(U_{\rm G}=0) $, as a function of $ V_{\rm D} $, becomes more prominent at low temperatures. \cite{Hasegawa_PRB04} The current-voltage characteristics deviate from those expected in the gradual channel approximation largely there. In fact, the assumption on which this approximation is based does not hold in the material with the quasi-one-dimensional and coherent band transport. This point will be discussed in the last section. At high temperatures, however, thermal fluctuations obscure the intrinsic nonlinearity and would make the characteristics similar to those expected in this approximation. 


In order to compare the field-effect characteristics in Mott insulators with those in band insulators, we used the tight-binding model with alternating transfer integrals, that with alternating site energies, and that with staggered magnetic fields. We show numerical results for the model with alternating transfer integrals only because the results for the other models are very similar. We use $ \delta t $=0.17, which also gives a charge gap of $ \Delta_{\rm CG} $=0.68 at half filling. 

\begin{figure}
\includegraphics[height=12cm]{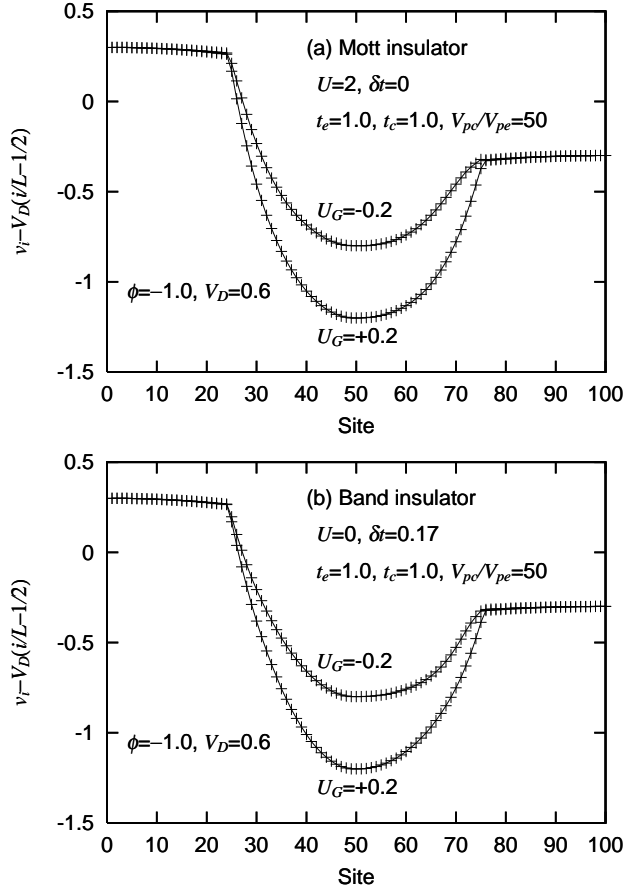}
\caption{Potential, $ v_i - V_{\rm D} (i/L-1/2) $, as a function of the site index, $ i $, at positive and negative gate biases $ U_{\rm G} $, (a) for the Mott insulator with $ U $=2 and $ \delta t $=0, and (b) for the band insulator with $ U $=0 and $ \delta t $=0.17. The crystal sites are $ 25 < i < 75 $. The other parameters are $ \phi $=$-$1.0, $ V_{\rm D} $=0.6, $ t_{\rm e} $=$ t_{\rm c} $=1, and $ V_{{\rm Pc}} / V_{{\rm Pe}} $=50.}
\label{fig:potential}
\end{figure}

Band insulators clearly show very asymmetric $ I_{\rm D} $-$ U_{\rm G} $ characteristics for finite work-function differences [Figs.~\ref{fig:band_IV}(b)- \ref{fig:band_IV}(d)]. Only when the work functions coincide, $ I_{\rm D} $ is suppressed only for $ \mid U_{\rm G} \mid < \Delta_{\rm CG}/2 $ and suddenly increases with $ U_{\rm G} $ or $ V_{\rm D} $ for $ \mid U_{\rm G} \mid > \Delta_{\rm CG}/2 $ [Fig.~\ref{fig:band_IV}(a)]. As shown later in Fig.~\ref{fig:potential}, the Schottky barriers with $ \phi < 0 $ are higher for the electron injections $ U_{\rm G} > 0 $ than for the hole injections $ U_{\rm G} < 0 $ at a given $ \mid U_{\rm G} \mid $. The drain current $ I_{\rm D} $ increases with the hole injections, but it remains suppressed for the electron injections. Although $ I_{\rm D} $ increases with $ V_{\rm D} $, this behavior is unchanged. A threshold in $ V_{\rm D} $ might be defined as long as we focus on $ I_{\rm D} $ near $ U_{\rm G}^{\rm min} $ only, but it is insignificant. These unipolar field-effect characteristics in band insulators are generally obtained irrespective of whether the charge gap is introduced by alternating transfer integrals, alternating site energies, or staggered magnetic fields. They are in marked contrast to the ambipolar field-effect characteristics in Mott insulators, where the charge gap is introduced by the electron correlation.

The asymmetric characteristics in band insulators with various work-function differences ($ \phi \neq 0 $) are similar if the polarity of the gate bias is fixed ($ \phi < 0 $ in the present case). With increasing the work-function difference $ \mid \phi \mid $, $ I_{\rm D} $ for the gate-bias polarity with higher Schottky barriers ($ U_{\rm G} > 0 $ in the present case) is further suppressed, while $ I_{\rm D} $ with lower Schottky barriers ($ U_{\rm G} < 0 $ in the present case) is further enhanced [Figs.~\ref{fig:band_IV}(b)-\ref{fig:band_IV}(d), see the vertical scales].

The spatial distribution of the total potential $ v_i - V_{\rm D} (i/L-1/2) $ is shown in Fig.~\ref{fig:potential}(a) for the Mott insulator, and in Fig.~\ref{fig:potential}(b) for the band insulator. Because the work function of the crystal is set lower than that of the electrodes, electrons transfer from the crystal to the electrodes, which make the second derivative of the potential positive in the crystal and negative in the electrodes, lowering the Fermi level of the crystal relative to that of the electrodes. For positive gate bias, $ U_{\rm G} > 0 $, the Fermi level of the crystal is further shifted downwards to accommodate electrons, so that the Schottky barriers at the interfaces are made higher. Meanwhile, for $ U_{\rm G} < 0 $, the potential is so modified as to accommodate holes, which make the Schottky barriers lower. In both cases, the potential has a long tail toward the middle of the crystal, which is in contrast to the MOSFET, where the larger impurity concentration shortens the depletion length. Roughly speaking, the potential distribution in the Mott insulator is similar to that in the band insulator. However, if one looks carefully at the potential difference between positive and negative gate biases, that for the Mott insulator is found different from that for the band insulator, especially near the crystal-electrode interfaces. 

\begin{figure}
\includegraphics[height=12cm]{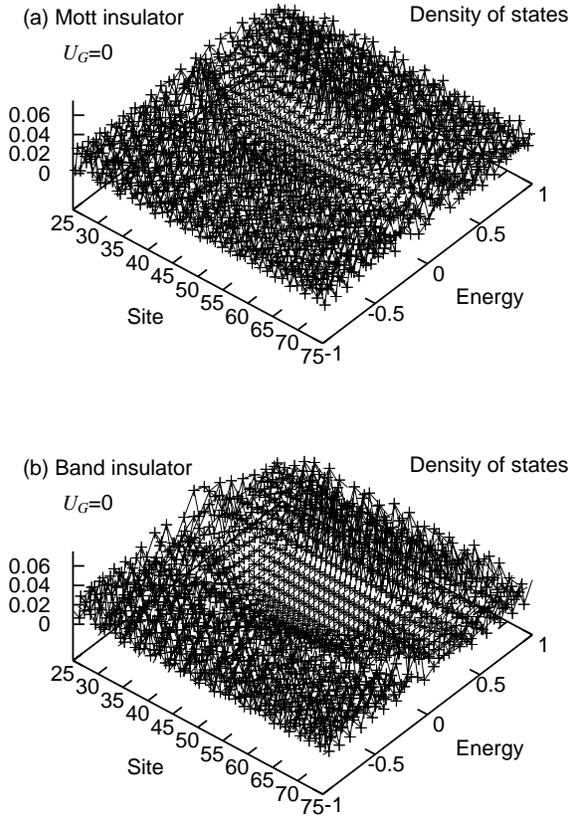}
\caption{Density of states in the crystal, $ 25 < i < 75 $, without gate bias, $ U_{\rm G} $=0, (a) for the Mott insulator with $ U $=2 and $ \delta t $=0, and (b) for the band insulator with $ U $=0 and $ \delta t $=0.17. The other parameters are the same as in Fig.~\ref{fig:potential}.}
\label{fig:DOS_UG00}
\end{figure}
In the case without gate bias ($ U_{\rm G} $=0), the spatial variation of the density of states in the crystal is shown in Fig.~\ref{fig:DOS_UG00}(a) for the Mott insulator, and in Fig.~\ref{fig:DOS_UG00}(b) for the band insulator. In both cases, the gap is formed, so that the drain current does not flow. The upper and lower bands correspond to the upper and lower Hubbard bands in Fig.~\ref{fig:DOS_UG00}(a), and to the conduction and valence bands in Fig.~\ref{fig:DOS_UG00}(b). At the bottom of the upper bands and at the top of the lower bands, the density of states increases abruptly, forming ridges. The ridges have a curvature according to the spatial variation of the potential shown in Fig.~\ref{fig:potential}. The space and energy dependence of the density of states in the Mott insulator is also similar to that in the band insulator as long as the drain current does not flow. Because their static properties are similar, the difference in their field-effect characteristics must be understood from the dynamical point of view. 

\begin{figure}
\includegraphics[height=12cm]{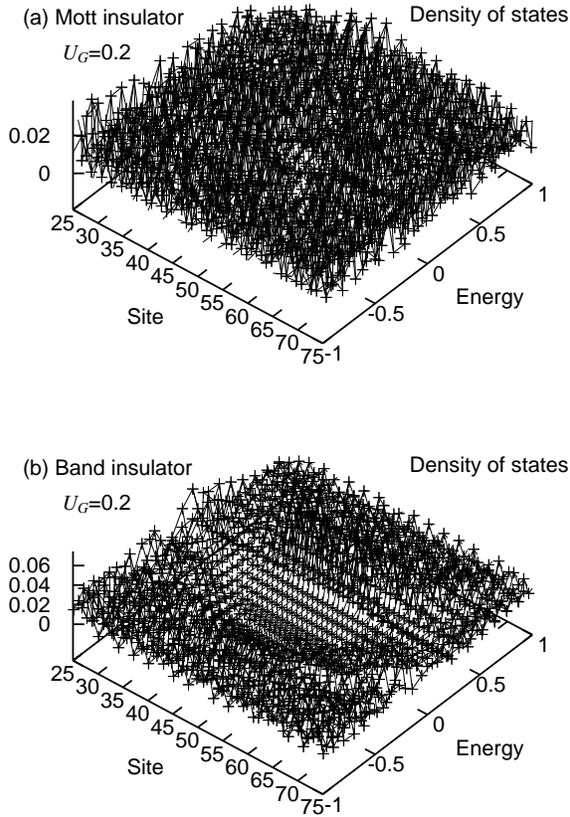}
\caption{Density of states in the crystal, $ 25 < i < 75 $, with a gate bias, $ U_{\rm G} $=0.2, (a) for the Mott insulator with $ U $=2 and $ \delta t $=0, and (b) for the band insulator with $ U $=0 and $ \delta t $=0.17. The other parameters are the same as in Fig.~\ref{fig:potential}.}
\label{fig:DOS_UG02}
\end{figure}
In the case with a finite gate bias ($ U_{\rm G} $=0.2), the spatial variation of the density of states in the crystal is shown in Fig.~\ref{fig:DOS_UG02}(a) for the Mott insulator, and in Fig.~\ref{fig:DOS_UG02}(b) for the band insulator. Here, the gap is formed in the band insulator only. The parameters used in Figs.~\ref{fig:DOS_UG02}(a) and \ref{fig:DOS_UG02}(b) here are the same as those in Figs.~\ref{fig:Mott_IV}(c) and \ref{fig:band_IV}(c), respectively, with $ V_{\rm D} $=0.6. The drain current flows in the Mott insulator here. Precisely speaking, the Hubbard system with the finite gate bias and the large-enough drain voltage is conducting. Thus the density of states for the Mott insulator becomes rather flat and featureless. The dynamic properties are determined in a self-consistent manner from the electron correlation and the Schottky barriers. This point will be discussed later.

\begin{figure}
\includegraphics[height=12cm]{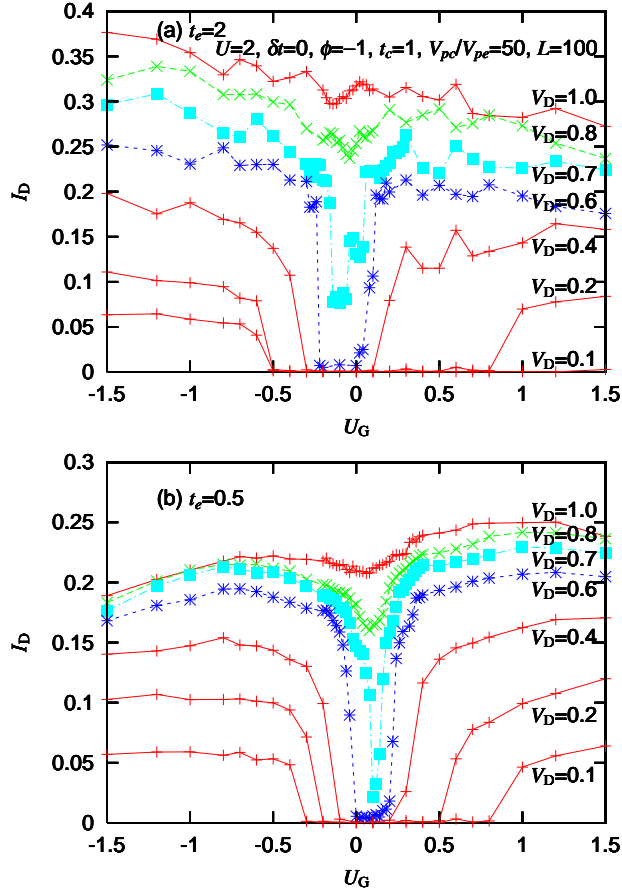}
\caption{$ I_{\rm D} $-$ U_{\rm G} $ characteristics at various drain voltage $ V_{\rm D} $ for Mott insulators with $ U $=2 and $ \delta t $=0. The bandwidth ratios are, (a) $ t_{\rm e} / t_{\rm c} $=2, and (b) $ t_{\rm e} / t_{\rm c} $=0.5. The other parameters are $ \phi $=$-$1.0, $ t_{\rm c} $=1, $ V_{{\rm Pc}} / V_{{\rm Pe}} $=50, and $ L $=100.}
\label{fig:bandwidth_ratio}
\end{figure}
In order to see how universal the field-effect characteristics of the Mott insulators are, we used different parameters. They have turned out to be always ambipolar. In Fig.~\ref{fig:bandwidth_ratio}(a), the bandwidth of the electrodes is set larger than that of the crystal. Compared with the figures already shown, the drain current shows rather irregular behavior especially for large $ V_{\rm D} $. This is due to numerically slower and poorer convergence generally for large bandwidths of the electrodes. Nevertheless, it is clear that the maximum of $ I_{\rm D} $ as a function of $ U_{\rm G} $ for large $ V_{\rm D} $ in Fig.~\ref{fig:Mott_IV}(c) disappears in the present case from the $ U_{\rm G} $ range shown here. In other words, the density of states in the electrodes substantially influences the behavior of $ I_{\rm D} $ as a function of $ U_{\rm G} $ for large $ V_{\rm D} $. Compared with Fig.~\ref{fig:Mott_IV}(c) for $ t_{\rm e} / t_{\rm c} $=1, the drain current is generally reduced by the larger $ t_{\rm e} $ here. It indicates that the mismatch of the bandwidths would contribute to backward scatterings of electrons entering the crystal, reducing the current flowing in the crystal. In Fig.~\ref{fig:bandwidth_ratio}(b), the bandwidth of the electrodes is set smaller than that of the crystal. Here again the drain current is reduced, compared with the case of $ t_{\rm e} / t_{\rm c} $=1. Obvious differences between Figs.~\ref{fig:bandwidth_ratio}(a) and \ref{fig:bandwidth_ratio} (b) are found in the deviations from the symmetric $ I_{\rm D} $-$ U_{\rm G} $ characteristics above $ V_{\rm Dth} $. In the former the hole injections are more effective in increasing $ I_{\rm D} $, while in the latter the electron injections are more effective. The minimum points $ U_{\rm G}^{\rm min} $ around $ V_{\rm Dth} $ are also shifted in the opposite directions from the point at $ t_{\rm e} / t_{\rm c} $=1.

\begin{figure}
\includegraphics[height=6cm]{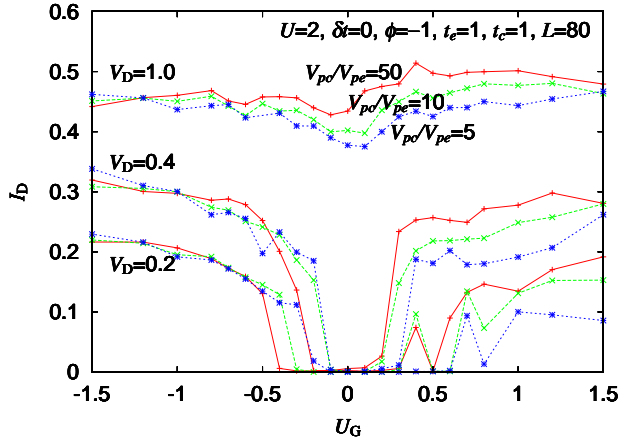}
\caption{$ I_{\rm D} $-$ U_{\rm G} $ characteristics at various drain voltage $ V_{\rm D} $ for Mott insulators with $ U $=2 and $ \delta t $=0. Different ratios of the Coulomb parameter in the crystal to that in the electrodes, $ V_{{\rm Pc}} / V_{{\rm Pe}} $, are used for comparison. The other parameters are $ \phi $=$-$1, $ t_{\rm e} $=$ t_{\rm c} $=1 with smaller $ L $=80 for convergence reasons.}
\label{fig:dielectric_ratio}
\end{figure}
In Fig.~\ref{fig:dielectric_ratio}, the ratio of the Coulomb parameters, $ V_{{\rm Pc}} / V_{{\rm Pe}} $, is varied. Because of numerically poorer convergence, we use a smaller system than before, but the drain current shows rather irregular behavior. With increasing the ratio $ V_{{\rm Pc}} / V_{{\rm Pe}} $, the $ I_{\rm D} $-$ U_{\rm G} $ characteristics appear to approach symmetric ones. In the realistic situation, this ratio would be much larger. In any case, the ambipolar field-effect characteristics are robust for the Mott insulators. We have shown the results with $ U $=2 for the Mott insulators. We checked the robustness by using different strengths of on-site repulsion. For instance, a weak on-site repulsion leads to a smaller charge gap, so that all the relevant energies including $ U_{\rm G} $ and $ V_{\rm D} $ are scaled down. The $ I_{\rm D} $-$ U_{\rm G} $ characteristics are always nearly symmetric below $ V_{\rm Dth} $, demonstrating the robustness.

\begin{figure}
\includegraphics[height=12cm]{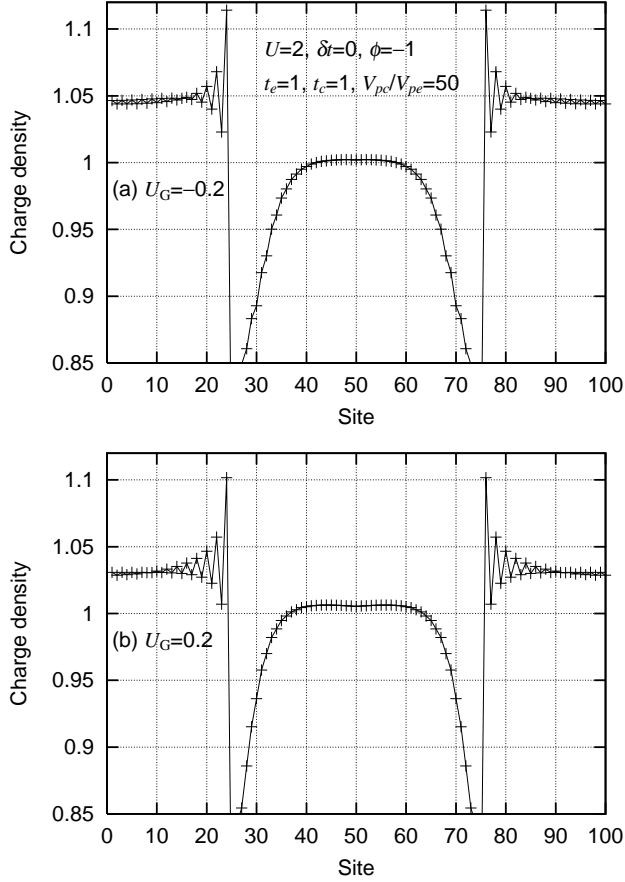}
\caption{Snapshots of charge density, (a) for $ U_{\rm G} $=$-$0.2, and (b) for $ U_{\rm G} $=0.2. The sites $ 25 < i < 75 $ are for the Mott insulator with $ U $=2 and $ \delta t $=0. The other parameters are $ \phi $=$-$1.0, $ t_{\rm e} $=$ t_{\rm c} $=1, and $ V_{{\rm Pc}} / V_{{\rm Pe}} $=50.}
\label{fig:snap_shot}
\end{figure}
In order to clarify the origin of the robustness of the ambipolar field-effect characteristics, we show snapshots of charge density ($ n_i $) for both gate bias polarities in Fig.~\ref{fig:snap_shot} before the drain voltage is applied. It is clearly seen that electrons transfer from the interfacial region inside the crystal to the whole region of the electrodes so as to adjust the relative Fermi levels when the work function of the crystal is set lower than that of the electrodes. Near the middle of the crystal, the charge density is almost flat and near unity due to the strong on-site repulsion.

For $ U_{\rm G} < 0 $ the Schottky barriers are lower [Fig.~\ref{fig:potential}(a), upper curve], and the depletion length is longer [Fig.~\ref{fig:snap_shot}(a)], while for $ U_{\rm G} > 0 $ the Schottky barriers are higher [Fig.~\ref{fig:potential}(a), lower curve] and the depletion length is shorter [Fig.~\ref{fig:snap_shot}(b)]. For the gate-bias polarity with the higher Schottky barriers ($ U_{\rm G} > 0 $), the deviation of the charge density well inside the insulator from unity is always larger than that for the opposite gate-bias polarity, indicating that the umklapp scattering is weaker. Therefore, the robustness of the ambipolar field-effect characteristics is a consequence of balancing the correlation effect with the barrier effect in a wide parameter space.

\begin{figure}
\includegraphics[height=12cm]{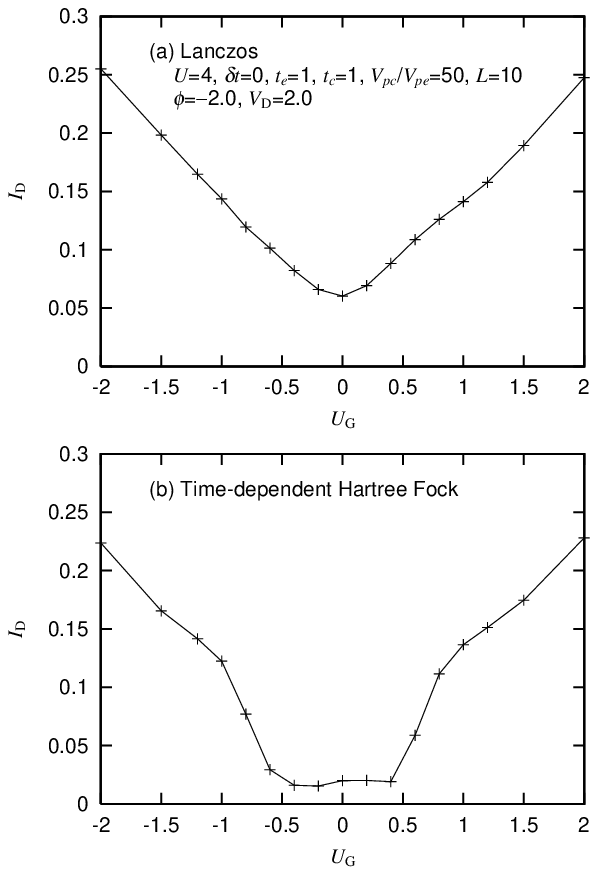}
\caption{$ I_{\rm D} $-$ U_{\rm G} $ characteristics for Mott insulators with $ U $=4 and $ \delta t $=0, (a) calculated with the help of the Lanczos method, and (b) calculated in the unrestricted Hartree-Fock approximation. The parameters are $ \phi $=$-$2.0, $ V_{\rm D} $=2.0, $ t_{\rm e} $=$ t_{\rm c} $=1, $ V_{{\rm Pc}} / V_{{\rm Pe}} $=50, and $ L $=10.}
\label{fig:lanczos_tdhf}
\end{figure}
So far, we have used the unrestricted Hartree-Fock approximation to replace the many-body wave function by the self-consistent Slater determinant of the one-body wave functions. In order to confirm that the ambipolar field-effect characteristics of the Mott insulators are not artifacts of this approximation, we employ the Lanczos method to treat the many-body wave function exactly. The potential distribution is obtained first in this approximation and averaged over the period, $ 0 < t < \Delta t $ with $ \Delta t = 2\pi L / ( 4 V_{\rm D}) $. With this static potential distribution, the time-dependent Schr\"odinger equation is solved by the Lanczos method. In the 10-site periodic system with 7 sites belonging to the crystal and the rest to the electrodes, 5 up-spin and 5 down-spin electrons are included. In this small system, a band insulator with a small gap shows rather symmetric $ I_{\rm D} $-$ U_{\rm G} $ characteristics if the localization length is comparable with or larger than the system size. Then, we compare the Mott insulator with $ U $=4 and the band insulator with $ \delta t $=0.77, both of which have a large charge gap of about $ \Delta_{\rm CG} $=3.1, and use the work-function difference $ \phi $=$-$2. The band insulator shows very asymmetric $ I_{\rm D} $-$ U_{\rm G} $ characteristics again: $ I_{\rm D} $ for $ U_{\rm G} $=$-$2 is about four times as much as $ I_{\rm D} $ for $ U_{\rm G} $=2 when $ V_{\rm D} $=2 is applied (not shown). On the other hand, the Mott insulator shows nearly symmetric $ I_{\rm D} $-$ U_{\rm G} $ characteristics, both by the Lanczos method [Fig.~\ref{fig:lanczos_tdhf}(a)] and in the Hartree-Fock approximation [Fig.~\ref{fig:lanczos_tdhf}(b)]. Therefore, the ambipolar field-effect characteristics are the intrinsic properties of Mott insulators. The magnitude of the charge gap is about $ \Delta_{\rm CG} $=3.1, both by the exact diagonalization and in the present approximation. The origin of the difference between the $ U_{\rm G} $ dependence of the exact $ I_{\rm D} $ and that of the approximate one is unclear.

\section{Summary and Discussion}

In order to clarify the mechanism of the ambipolar field-effect characteristics in MISFET device structures based on organic single crystals of the quasi-one-dimensional Mott insulator (BEDT-TTF)(F$_2$TCNQ), \cite{Hasegawa_PRB04} we have performed model calculations taking account of the fact that the gate bias does not only modulate the carrier density but also affect the Schottky barriers in this low-dimensional system showing coherent band transport. \cite{Hasegawa_PRB04} The one-dimensional Hubbard model is used for Mott insulators and attached to the tight-binding model for source and drain electrodes with a generally different work function. The scalar and vector potentials satisfying the Poisson equation are added to the model, which originate from the long-range Coulomb interaction and form the Schottky barriers. The drain voltage and the gate bias modulate the boundary values of the Poisson equation. The time-dependent Schr\"odinger equation is solved in the unrestricted Hartree-Fock approximation, simultaneously with the Poisson equation. The self-consistency is imposed at each site and time in both equations.

Mott insulators show ambipolar field-effect characteristics in a very wide parameter space spanned by the work-function difference, the bandwidth difference, and the Coulomb parameters in the Poisson equation. Thus, the experimentally observed, ambipolar carrier injections are indeed general characteristics of quasi-one-dimensional Mott insulators. For the gate-bias polarity with higher Schottky barriers, the correlation effect is weakened accordingly. In other words, the robust ambipolar characteristics are caused by balancing the correlation effect with the barrier effect. This is in marked contrast to the fact that band insulators generally show unipolar field-effect characteristics as long as the work function of the crystal is different from that of the electrodes.

The difference in the transport property can be understood as due to the difference in the collectivity. To contribute to the current density, carriers must go through the crystal-electrode interfaces. When the crystal is a band insulator, each carrier is affected by the Schottky barrier independently of other carriers. As the barrier becomes higher, it becomes more difficult for the carrier to go through it, so that the current density is reduced. When the crystal is a Mott insulator, all the electrons are correlated, repelling nearby electrons. If the electron density deviates from the equilibrium one in such a way that the positive or negative charge density is locally accumulated, it is energetically unfavorable. Thus the electrons move collectively. This is why the transport property is insensitive to the local variation of the potential distribution near the interfaces in Mott insulators. The charge density distribution is determined in a self-consistent manner to satisfy both the Schr\"odinger equation and the Poisson equation. Thus the influence of the on-site repulsion and that of the long-range Coulomb interaction counterbalance each other. Balancing the correlation effect with the barrier effect causes the robust ambipolar characteristics in Mott insulators.

Such collective transport is reminiscent of the knock-on mechanism in ion channels. \cite{Roux_Nature,Roux_PNAS} K$^+$ ions hop in single file (i.e., in a queue) from one binding site to the next as permeation proceeds. Although hopping processes occur temporally in a random fashion, they are spatially correlated. The neighboring ions hop together. Roughly speaking, the occupation of ions in the binding sites is either 0101 or 1010, which is reminiscent of a charge-ordered Mott insulator. The effective electrostatic repulsion between the ions is manifested at short distances. The concerted multi-ion conduction mechanism is determined largely by the multi-ion free energy surface rather than the dissipative and frictional forces. Thus the ion translocation in the multi-ion potential energy surface is essential for achieving a high throughput rate, as clarified by a modified Poisson-Boltzmann theory. \cite{Roux_Nature,Roux_PNAS}

When the motion of carriers is coherent and restricted to a one-dimensional system, the gradual channel approximation does not generally work because the contact resistance dominates the channel conductance. Thus the barrier effect is relatively important. In this sense, the mobility reported in the experiment on the quasi-one-dimensional Mott insulator, \cite{Hasegawa_PRB04} is an effective one because the mobility is estimated by fitting the current-voltage relation to that based on the gradual channel approximation. The fact that the mobility increases with the drain voltage and depends on the polarity of the gate bias indicates that it is an effective one. The estimated mobility values are indeed very low (by some three orders of magnitude), although the coherent band transport is expected from the temperature dependence of the mobility. \cite{Hasegawa_PRB04} The coherent band transport would enhance the relative importance of the Schottky barriers.

It is true that the intermolecular overlaps along the molecular long axis are finite, although they are very small. Considering the transverse field (i.e., the transverse distribution of the potential) and the consequent transverse motion of electrons are necessary not only when a crossover from one-dimensional to three-dimensional systems is discussed but also when the gate bias is so large that the modulation of the longitudinal potential distribution, especially near the crystal-(source/drain) electrode interfaces, is not described by the gate efficiency factor.

\section*{Acknowledgment}

The author is grateful to T. Hasegawa for showing his data prior to publication and for enlightening discussions. 
He thanks N. Maeshima for computational support with the time-dependent Lanczos method. 
This work was supported by Grants-in-Aid for Scientific Research (C) (No. 15540354), for Scientific Research on Priority Area ``Molecular Conductors'' (No. 15073224), for Creative Scientific Research (No. 15GS0216), and NAREGI Nanoscience Project from the Ministry of Education, Culture, Sports, Science and Technology, Japan.

\appendix
\section{Gauge Invariance of Electron Dynamics}

For simplicity, we treat the one-dimensional model containing the scalar-potential and transfer terms only, by setting $ \epsilon_i = U_i = 0 $ and $ \mid t_{i,i+1}(t) \mid = t_0 $ in eq.~(\ref{eq:Hamiltonian}). The total force from the scalar and vector potentials is shown to be given symbolically by 
\begin{equation}
-e E_{\rm tot} = -e E_{\rm D} - \partial_x v 
\;,
\end{equation}
independently of the gauge. For intuitive understanding, let us first take a plane wave, $ \exp[ {\rm i}kr ] $ at the position $ r=ja $ with integer $ j $. The Peierls phase, $ \exp \left[ {\rm i} ea A(t)/(\hbar c) \right] = \exp \left[ -{\rm i} e E_{\rm D} a t/\hbar \right] $, shifts its momentum $ \hbar k $ by $ -e E_{\rm D} t $, which corresponds to the force $ -e E_{\rm D} $ from the vector potential. Thus the above equation looks reasonable. 

To prove the above equation, we consider how the expectation value of a (generally off-diagonal) one-body density, $ \langle \exp \left( {\rm i} n e E_{\rm D} a t / \hbar \right) c_{j+n,\sigma}^\dagger c_{j,\sigma} \rangle $, evolves in time. It is straightforward to derive the equation,
\begin{eqnarray}
& & {\rm i} \hbar \partial_t 
\langle   \exp \left[ {\rm i} n e E_{\rm D} a t / \hbar \right] 
c_{j+n,\sigma}^\dagger c_{j,\sigma}   \rangle   \nonumber \\
& & = \left[  - n e E_{\rm D} a - \left( v_{j+n} -v_{j} \right )  \right]
\exp \left[ {\rm i} n e E_{\rm D} a t / \hbar \right] 
\langle c_{j+n,\sigma}^\dagger c_{j,\sigma} \rangle   \nonumber \\
& & - t_0 \exp \left[ {\rm i} (n-1) e E_{\rm D} a t / \hbar \right] 
\langle c_{j+n,\sigma}^\dagger c_{j+1,\sigma} 
- c_{j+n-1,\sigma}^\dagger c_{j,\sigma} \rangle   \nonumber \\
& & + t_0 \exp \left[ {\rm i} (n+1) e E_{\rm D} a t /\hbar \right] 
\langle c_{j+n+1,\sigma}^\dagger c_{j,\sigma}
- c_{j+n,\sigma}^\dagger c_{j-1,\sigma} \rangle   
\;.
\end{eqnarray}
This shows that the time-derivative of $ \langle \exp \left( {\rm i} n e E_{\rm D} a t / \hbar \right) c_{j+n,\sigma}^\dagger c_{j,\sigma} \rangle $ for any integer $ n $ is always given by a linear combination of themselves for $ n = 0, 1, \cdots, L-1 $. All the coefficients are either $ \pm t_0 $ or $ \left[  - n e E_{\rm D} a - \left( v_{j+n} -v_{j} \right )  \right] $. Thus their time-evolution is governed by the contribution from the vector potential, $ - n e E_{\rm D} a $, and that from the scalar potential, $ - \left( v_{j+n} -v_{j} \right ) $, which appear always as the sum of them. Symbolically, it corresponds to $ n a $ times the quantity $ -e E_{\rm D} - \partial_x v $ in the continuum limit. Thus the above equation is proved. 

This proof is easily extended to the case, in which the vector potential depends on the bond on which a finite transfer integral is assigned. The time-derivative of $ \langle \exp \left( {\rm i} e ( E_{{\rm D}j+n-1/2} + \cdots + E_{{\rm D}j+1/2} ) a t / \hbar \right) c_{j+n,\sigma}^\dagger c_{j,\sigma} \rangle $ for any integer $ n $ is again given by a linear combination of themselves for $ n = 0, 1, \cdots, L-1 $. All the coefficients are then either $ \pm t_0 $ or $ \left[  - e ( E_{{\rm D}j+n-1/2} + \cdots + E_{{\rm D}j+1/2} ) a - \left( v_{j+n} -v_{j} \right )  \right] $. If we take a space-dependent vector potential, it modifies the Poisson equation. Thus the convenient gauge is the space-independent vector potential that absorbs the finite $ V_{\rm D} $ so that the scalar potential is periodic and the vector potential is not involved in the Poisson equation.


\begin{thebibliography}{99} 
\bibitem{Sze_book} S. M. Sze: {\it Physics of Semiconductor Devices} (John Wiley \& Sons, New York, 1981) 2nd ed.
\bibitem{Heinze_PRL} S. Heinze, J. Tersoff, R. Martel, V. Derycke, J. Appenzeller and Ph. Avouris: Phys. Rev. Lett. \textbf{89} (2002) 106801.
\bibitem{Appen_PRL02} J. Appenzeller, J. Knoch, V. Derycke, R. Martel, S. Wind and Ph. Avouris: Phys. Rev. Lett. \textbf{89} (2002) 126801.
\bibitem{Rado_APL03} M. Radosavljevi\'c, S. Heinze, J. Tersoff and Ph. Avouris: Appl. Phys. Lett. \textbf{83} (2003) 2435.
\bibitem{Javey_Nature} A. Javey, J. Guo, Q. Wang, M. Lundstrom and H. Dai: Nature \textbf{424} (2003) 654.
\bibitem{Kishine_IJMPB} J. Kishine and K. Yonemitsu: Int. J. Mod. Phys. B \textbf{16} (2002) 711.
\bibitem{Hasegawa_PRB04} T. Hasegawa, K. Mattenberger, J. Takeya and B. Batlogg: Phys. Rev. B \textbf{69} (2004) 245115.
\bibitem{Hasegawa_PRB00} T. Hasegawa, T. Mochida, R. Kondo, S. Kagoshima, Y. Iwasa, T. Akutagawa, T. Nakamura and G. Saito: Phys. Rev. B \textbf{62} (2000) 10059.
\bibitem{Roux_Nature} S. Bern\`eche and B. Roux: Nature \textbf{414} (2001) 73.
\bibitem{Roux_PNAS} S. Bern\`eche and B. Roux: Proc. Natl. Acad. Sci. USA \textbf{100} (2003) 8644.
\bibitem{Javey_APL02} A. Javey, M. Shim and H. Dai: Appl. Phys. Lett. \textbf{80} (2002) 1064.
\bibitem{Miyashita_JPSJ03} N. Miyashita, M. Kuwabara and K. Yonemitsu: J. Phys. Soc. Jpn. \textbf{72} (2003) 2282.
\bibitem{Oka_PRL03} T. Oka, R. Arita and H. Aoki: Phys. Rev. Lett. \textbf{91} (2003) 066406.
\bibitem{Hasegawa_SSC97} T. Hasegawa, S. Kagoshima, T. Mochida, S. Sugiura and Y. Iwasa: Solid State Commun. \textbf{103} (1997) 489.
\end{thebibliography}
\end{document}